# Correlated charged impurity scattering in graphene

Jun Yan and Michael S. Fuhrer*

*Center for Nanophysics and Advanced Materials and Materials Research Science and Engineering Center, University of Maryland, College Park, MD 20742, USA*

\* Email: mfuhrer@umd.edu

**Understanding disorder in graphene is essential for electronic applications; in contrast to conventional materials, the extraordinarily low electron-phonon scattering[1,2] in graphene implies that disorder[3-7] dominates its resistivity even at room temperature. Charged impurities[5,8-10] have been identified as an important disorder type in graphene on $SiO_2$ substrates[11,12], giving a nearly linear carrier-density-dependent conductivity $\sigma(n)$, and producing electron and hole puddles[13-15] which determine the magnitude of graphene's minimum conductivity $\sigma_{min}$[10]. Correlations of charged impurities are known to be essential in achieving the highest mobilities in remotely-doped semiconductor heterostructures[16-18], and are present to some degree in any impurity system at finite temperature. Here we show that even modest correlations in the position of charged impurities, realized by annealing potassium on graphene, can increase the mobility by more than a factor of four. The results are well understood theoretically[19] considering an impurity correlation length which is temperature dependent but independent of impurity density. Impurity correlations also naturally explain the sub-linear $\sigma(n)$ commonly observed in substrate-bound graphene devices[2,11,12,20].**

Our experiment probes the influence of thermal annealing on the electronic transport properties of a graphene device with adsorbed potassium (K) atoms; potassium donates an electron to graphene leaving a positive ion as a charged scattering center. We previously studied[21] charged impurity scattering in potassium/graphene at low temperature (20 K) where the potassium adatoms are presumed to be frozen randomly on the graphene. Here we measure the transport properties of graphene as the temperature is raised. We expect potassium ions on graphene to experience mutual Coulomb repulsion which drives them away from each other producing correlations in their positions. Indeed, early low-energy electron diffraction (LEED) studies of potassium adsorbed on graphite revealed a distinct diffraction peak, which is linked to the nearest-neighbor spacing of the dispersed potassium layer[22]. A recent scanning tunneling microscope (STM) study reached similar conclusions[23].

The graphene on $SiO_2$/Si sample was prepared by mechanical exfoliation, similar to our previous work[1] (see Methods). Figure 1 (left inset) shows a photomicrograph of the device. Figure 1 (right inset) shows the conductivity $\sigma$ of the pristine device as a function of electron density $n = c_g(V_g - V_{g,min})/e$ where $V_g$ is the gate voltage, $V_{g,min}$ is the gate voltage of minimum conductivity, $c_g = 11$ nF/cm$^2$ is the gate capacitance, and $e$ is the elementary charge. The field effect mobility of the sample $\mu_{FE} = \frac{1}{e}\frac{d\sigma}{dn}$ is about 20,000 cm$^2$/Vs, which is among the highest for graphene deposited on $SiO_2$/Si substrates[11]. Assuming Matthiessen's rule for long range and short range scatters the transport curve can be quantitatively described (red curve) by

$$\sigma_{pristine}(n) = \left(\frac{1}{ne\mu_L} + \rho_S\right)^{-1} \quad (1)$$

(see Supplementary Materials for details of the fitting procedure). Equation (1) has been interpreted as reflecting scattering by uncorrelated charged impurities ($\mu_L$) and weak point disorder ($\rho_S$)[2, 11, 12, 20].

Potassium was deposited on the sample at low temperature $T = 20$ K then $\sigma(n)$ was measured at various temperatures from 20 K to 180 K. The sample was then baked at high temperatures to remove residual adsorbates, cooled to low temperature, and the experiment was repeated with a different potassium density. We observed no degradation of sample quality upon repeated experiments; the sample mobility varied by ±10% and the $\sigma_{min}$ occured at gate voltages -1V $\leq V_{g,min} \leq$ 1V (see Supplementary Materials).

Figure 1 (main panel) summarizes the effects of increasing potassium density on $\sigma(V_g)$ at $T = 20$ K. $V_{g,min}$ shifts to increasingly negative gate voltages with increased potassium density, reflecting electron doping by potassium. The sample mobility decreases by more than an order of magnitude. These observations agree well with our previous studies[21].

Figure 2a shows $\sigma(V_g)$ at different temperatures for a potassium doping level that results in a $V_{g,min}$ shift $\Delta V_{g,min} \approx 62$ V. The conductivity increases with temperature, more rapidly for $T > 100$K. The minimum conductivity point $V_{g,min}$ remains fixed for $T < 180$ K, indicating that doping by potassium persists. (For T > 180 K, we observe $V_{g,min}$ shifts toward 0 V, indicating potassium migration off the sample or desorption.) In addition to mobility improvements, $\sigma(V_g)$ also becomes significantly sub-linear at elevated temperatures, in contrast to the linear $\sigma(V_g)$ expected[5, 8-10, 24] and observed[21, 25] for isolated or clustered charged impurity scattering, and observed here at $T = 20$ K. Figure 2b shows the maximum field effect mobility as a function of temperature for various potassium densities reflected in different $\Delta V_{g,min}$. The field effect mobility increases over fourfold for the largest $\Delta V_{g,min}$.

In the Boltzmann formalism for charge transport, the square of screened Coulomb scattering potential $|\tilde{V}(q)|^2$ enters the relaxation time approximation when the charged impurities are uncorrelated. In the presence of correlation, estimation of relaxation time needs to take into account the structure factor $S(q)$ of the scattering centers and $|\tilde{V}(q)|^2$ is replaced with $|\tilde{V}(q)|^2 S(q)$ [26]. The structure factor is linked to the spatial distribution of potassium ions via a Fourier transformation. Here we model the spatial correlation with a simple pair distribution function $g(r)$ recently proposed by Li *et al.* in Ref.19: $g(r)$ is 0 for $r < r_c$ and 1 for $r > r_c$ where $r_c$ is the correlation length, the single additional fit parameter. The corresponding structure factor is

$$S(q) = 1 - 2\pi n_K \frac{r_c}{q} J_1(qr_c) \qquad (2)$$

where $J_1$ is the Bessel function of the first kind, and $n_K$ is the potassium density. The resistivity $\rho_K(n, n_K, r_c)$ due to scattering by correlated potassium ions may then be calculated by numerical integration (see Methods).

Taking further into account the fact that there is some initial disorder in pristine graphene (right inset of Fig.1) and that there exists temperature-dependent acoustic phonon scattering[1], we fit our transport curves with the following expression:

$$\sigma(n,n_K,T) = \left(\sigma_{pristine}(n)^{-1} + \rho_{ph}(T) + \rho_K(n,n_K,r_c(T))\right)^{-1}. \quad (3)$$

$\sigma_{pristine}(n)$ is determined by fitting to equation (1) (see Fig. 1 right inset) and $\rho_{ph}$ = [0.1 Ω/K]×$T$. The only free fitting parameters in equation (3) are $n_K$ and $r_c$.

For each set of data we treat $n_K$ as a global parameter while $r_c$ varies with temperature. We note that $r_c$ cannot be smaller than 4.9Å since the most dense potassium overlayer on graphene is the close packed 2×2 ($C_8K$) structure[22]. In our fits we fix $r_c$ = 4.9Å at base temperature. With these considerations we find that our data are well described by equation (3) as shown by the dashed lines in Fig. 2a and Fig. 3 (three additional data sets as well as fits are shown in Supplementary Materials). The fits not only describe the mobility increase but also capture the increase in the curvature in σ($n$).

Figure 4 summarizes the fit parameters. The correlation length is found to increase monotonically with temperature, and is insensitive to potassium density which varies over an order of magnitude; this scaling of the correlation length lends support to its physicality. The correlation lengths found in Fig.4 are smaller than the K-K distances $\pi n_K r_c^2 < 1$ even at the highest temperatures, consistent with this regime where the pair distribution model[19] is applicable. Using the convention that the close packed 2×2 potassium overlayer ($C_8K$) corresponds to the coverage $\theta$=1, the regime that is studied here is 0.001< $\theta$ <0.01. At similar potassium coverage, LEED studies reveal that the potassium overlayer on graphite gives rise to a distinct diffraction peak that moves to higher wavevectors with adding of potassium and becomes better defined at higher temperatures[27]. These observations are in accord with our experimental results, further substantiating our interpretation that correlation between potassium ions improves with temperature and strongly influences the transport properties of graphene devices. A recent study reported[23] strongly correlated potassium ($\pi n_K r_c^2 \approx 1$) deposited on graphite at 11 K at a density about twice the highest density studied here, probably reflecting a much more disordered landscape for potassium on highly-corrugated[28] graphene on $SiO_2$.

Figure 4 inset shows $n_K$ as a function of $\Delta V_{g,min}$. At high potassium densities (greater than the initial impurity density $n_{imp}$ ~ 4 × $10^{11}$ cm$^{-2}$; see below), the experimentally extracted $n_K$ vs. $\Delta V_{g,min}$ can be described by the self-consistent theory for graphene in the presence of random charged impurity disorder[10, 21] with the fitting parameter $d$ (distance of impurity to the graphene plane) equal to 1 nm. This behavior indicates incomplete screening by graphene[10]. At low potassium densities $n_K < n_{imp}$ ~ 4 × $10^{11}$ cm$^{-2}$, there is no theoretical prediction, but the simple prediction from geometric capacitance $n_K = \dfrac{C_g \Delta V_{g,min}}{e}$ (red line) describes our data well.

Figure 5 shows the temperature dependence of $\sigma_{min}$. $\sigma_{min}$ varies only slightly with potassium doping, as previously observed[21]. Within the self-consistent theory[10], $\sigma_{min}$ = $n^* e \mu$, where $n^*$ is the residual (puddle) carrier density, and $\mu$ the mobility. Interestingly, the temperature dependence of $\sigma_{min}$ is weak, and is very similar to the undoped case ($\Delta V_{g,min}$ = 0). This is surprising, given the large increase in mobility; it suggests a large decrease in the effective (puddle) carrier density at the minimum point for correlated disorder. More work is needed to understand this behavior.

The fact that impurity correlations always produce sublinear $\sigma(n)$ prompts us to revisit the interpretation of equation (1). While experimental evidence for long range scattering ($\mu_L$) prevails[13-15, 21], sources for the short range scattering ($\rho_s$) are mysterious and so far have not been well established. In particular STM measurements have found that point defects in graphene lattice are extremely rare[14, 15]. Meanwhile, it is quite likely that the long range scatters are correlated to some degree. We find that correlations in long-range scatterers alone can explain the observed sub-linearity in $\sigma(n)$ without invoking point disorder.

In the right inset of Fig.1 we refit $\sigma(n)$ for the pristine graphene sample to the theory for correlated charged impurities, and the result is shown as the blue curve; fit parameters are impurity density $n_{imp}$ = 4.6 (3.9) ×10$^{11}$ cm$^{-2}$ and $r_c$ = 6.1 (7.0) nm for electrons (holes). The fit is almost indistinguishable from equation (1). This is not surprising; for small argument of the Bessel function in equation (2) i.e. $\pi n r_c^2 \lesssim 1$, $\sigma(n)$ is well approximated[19] by equation (1), with $\mu_L = \mu_0/(1 - \alpha)$ and $\rho_s$ = 290 Ω × $\alpha^2$ where $\mu_0$ is the mobility for uncorrelated charged impurities, $\alpha = \pi n_{imp} r_c^2 < 1$ (see Methods). This is consistent with the range of observed $\rho_s$ of 50-100 Ω on SiO$_2$[2, 12] and h-BN[20]. Charges are known to be mobile on the surface of SiO$_2$ on a timescale of seconds at room temperature[29]. Assuming that the SiO$_2$ mobile surface charges correspond to a nondegenerate plasma frozen at a temperature $T_0$[16], the correlation length $r_c = \kappa k_B T_0/n_{imp}e^2 \sim 6$ nm predicts $T_0 \sim$ 170 K which is a plausible temperature for freezing the oxide trapped charge configuration. More experiments are needed to understand the degree of correlation of disorder in various substrates used for graphene devices, but intentional correlation of disorder e.g. by control of charge trap distributions or by rapid thermal annealing and quenching should be a powerful tool to increase mobility in graphene devices.

**Methods**

The monolayer graphene sample is mechanically exfoliated from natural graphite (Nacional de Grafite Ltda.) and deposited on a heavily doped silicon substrate with 300nm thick $SiO_2$. Electrical contacts are defined with standard electron beam lithography using poly(methyl methacrylate) resist and thermal evaporation of chromium/gold. An additional electron beam lithography step defines a mask through which the sample is then etched with oxygen plasma into the shape as shown in the left inset of Fig.1. To remove surface contamination after the sample fabrication process, the sample is annealed in $H_2$/Ar gas at 350°C for an hour[30] before mounting it on a cold finger inside an ultra-high vacuum (UHV) chamber with a base pressure of $6\times10^{-10}$ Torr. Transport properties of the device are measured using a standard four-probe measurement configuration. The sample is found to be p-type (hole-doped) at this point. The sample and UHV chamber are baked at 200°C under vacuum for a few days to further improve surface cleanliness; after baking the sample is found to be nearly undoped. Potassium evaporation is achieved by electrically heating up a getter source (SAES Getters).

The scattering time for correlated charged impurities is calculated from Fermi's golden rule: $\frac{\hbar}{\tau} = \frac{n_K}{4\pi} \int d^2k \left(\tilde{V}(q)\right)^2 S(q) [1 - \cos^2\theta] \delta(E_k - E_F)$, where $q = 2k_F \sin(\theta/2)$ is the momentum transferred by scattering, with the Fermi wavevector $k_F = \sqrt{\pi n}$; the screened Coulomb potential $\tilde{V}(q) = \frac{V(q)}{\varepsilon(q)}$; the bare Coulomb potential $V(q) = 2\pi e^{-qd} e^2/\kappa q \approx 2\pi e^2/\kappa q$ for $k_F d \ll 1$, where $d$ is the impurity-graphene separation and $\kappa$ the average dielectric constant of substrate and vacuum; the dielectric function of graphene $\varepsilon(q) = 1 + 4k_F r_s/q$; $r_s = e^2/\hbar v_F \kappa$ is graphene's fine structure constant; $S(q)$ is the structure factor given by equation (2) [19]. We then have $\frac{\hbar}{\tau} = \left(\frac{\pi n_K \hbar v_F r_s^2}{4k_F}\right) \int d\theta \frac{(1 - \cos^2\theta)}{\left(\sin\frac{\theta}{2} + 2r_s\right)^2} S\left(2k_F \sin\frac{\theta}{2}\right)$ which is solved by numerical integration to give $\tau$. The resistivity is given by $\rho_K = \left(\frac{h}{e^2}\right) \frac{\hbar}{2E_F \tau}$.

For $k_F r_c \lesssim 1$, the structure factor in equation (2) is approximated by $S(q) = 1 - \pi n_{imp} r_c^2 (1 - q^2 r_c^2/8)$. Then the resistivity may be calculated analytically and cast into the same format as that of equation (1) with the parameters given by $\mu_L = \mu_0/(1 - \alpha)$ and $\rho_s = (h/e^2)[r_s^2 G_2(r_s)\alpha^2]$, where $\mu_0$ is the mobility for a density $n_{imp}$ of uncorrelated charged impurities[8], $G_2(x) = \frac{\pi}{16} - \frac{4x}{3} + 3\pi x^2 + 40x^3 \left[1 - \pi x + \frac{4}{5}(5x^2 - 1)\frac{\arccos(1/2x)}{\sqrt{4x^2 - 1}}\right]$. For SiO$_2$ and h-BN, $\kappa \approx 2.5$, $r_s \approx 0.8$, and $G_2(0.8) \approx 0.018$.

**REFERENCES**

1. Chen, J.-H., Jang, C., Xiao, S., Ishigami, M. & Fuhrer, M. S. Intrinsic and extrinsic performance limits of graphene devices on SiO$_2$. *Nature Nanotech.* **3,** 206–209 (2008).

2. Morozov, S. V. *et al*. Giant intrinsic carrier mobilities in graphene and its bilayer. *Phys. Rev. Lett.* **100,** 016602 (2008).

3. Shon, N. H. & Ando, T. Quantum transport in two-dimensional graphite system. *J. Phys. Soc. Jpn.* **67,** 2421-2429 (1998).

4. Peres, N. M. R., Guinea, F. & Castro Neto A. H. Electronic properties of disordered two-dimensional carbon. *Phys. Rev. B* **73,** 125411 (2006).

5. Ando, T. Screening effect and impurity scattering in monolayer graphene. *J. Phys. Soc. Jpn.* **75,** 074716 (2006).

6. Chen, J.-H., Cullen, W. G., Jang, C., Fuhrer, M. S. & Williams E. D. Defect scattering in graphene. *Phys. Rev. Lett.* **102,** 236805 (2009).

7. Robinson, J. P., Schomerus, H., Oroszlány, L. & Falko, V.I. Adsorbate-limited conductivity of graphene. *Phys. Rev. Lett.* **101,** 196803 (2008).

8. Hwang, E. H., Adam, S. & Das Sarma, S. Carrier transport in two-dimensional graphene layers. *Phys. Rev. Lett.* **98,** 186806 (2007).

9. Nomura, K. & MacDonald, A. H. Quantum transport of massless Dirac fermions. *Phys. Rev. Lett.* **98,** 076602 (2007).

10. Adam, S., Hwang, E. H., Galitski, V. M., & Das Sarma, S. A self-consistent theory for graphene transport. *Proc. Natl Acad. Sci. USA* **104,** 18392-18397 (2007).

11. Tan, Y.-W. *et al.* Measurement of scattering rate and minimum conductivity in graphene. *Phys. Rev. Lett.* **99,** 246803 (2007).

12. Jang, C., Adam, S., Chen, J.-H., Williams, E. D., Das Sarma, S. & Fuhrer, M. S. Tuning the effective fine structure constant in graphene: opposing effects of dielectric screening on short- and long-range potential scattering. *Phys. Rev. Lett.* **101,** 146805 (2008).

13. Martin, J. *et al.* Observation of electron–hole puddles in graphene using a scanning single-electron transistor. *Nature Phys.* **4,** 144-148 (2007).

14. Xue, J. *et al.* Scanning tunnelling microscopy and spectroscopy of ultra-flat graphene on hexagonal boron nitride. *Nature Mater.* **10,** 282–285 (2011).

15. Zhang, Y. *et al.* Origin of spatial charge inhomogeneity in graphene. *Nature Phys.* **5,** 722 - 726 (2009).

16. Efros, A. L., Pikus, F. G. & Samsonidze, G. G. Maximum low-temperature mobility of two-dimensional electrons in heterojunctions with a thick spacer layer. *Phys. Rev. B* **41,** 8295 (1990).

17. Coleridge, P. T. Small-angle scattering in two-dimensional electron gases. *Phys. Rev. B* **44,** 3793 (1991).

18. Buks, E., Heiblum, M. & Shtrikman, H. Correlated charged donors and strong mobility enhancement in a two-dimensional electron gas. *Phys. Rev. B* **49,** 14790 (1994).

19. Li, Q., Hwang, E. H., Rossi E. & Das Sarma, S. Theory of 2D transport in graphene for correlated disorder. arXiv:1104.0667 (2011).

20. Dean, C. R. *et al.* Boron nitride substrates for high-quality graphene electronics. *Nature Nanotech*. **5,** 722 (2010).

21. Chen, J.-H. *et al.* Charged-impurity scattering in graphene. *Nature Phys*. **4,** 377-381 (2008).

22. Li, Z. Y., Hock, K. M. & Palmer, R. E. Phase transitions and excitation spectrum of submonolayer potassium on graphite. *Phys. Rev. Lett.* **67,** 1562-1565 (1991).

23. Renard, J., Lundeberg, M. B., Folk, J. A. & Pennec, Y. Real-time imaging of K atoms on graphite: Interactions and Diffusion. *Phys. Rev. Lett.* **106,** 156101 (2011).

24. Katsnelson, M. I., Guinea, F. & Geim, A. K. Scattering of electrons in graphene by clusters of impurities. *Phys. Rev. B* **79**, 195426 (2009).

25. McCreary, K. M. *et al.* Effect of cluster formation on graphene mobility. *Phys. Rev. B* **81**, 115453 (2010).

26. Mahan, G. D. *Many-Particle Physics* (Plenum, New York, 1990).

27. Hock, K. M. & Palmer, R. E. Temperature dependent behaviour in the adsorption of submonolayer potassium on graphite. *Surf. Sci.* **284,** 349 (1993).

28. Cullen, W. *et al.* High-fidelity conformation of graphene to $SiO_2$ topographic features. *Phys. Rev. Lett.* **105,** 215504 (2010).

29. Lambert, J. *et al.* Dispersive charge transport along the surface of an insulating layer observed by electrostatic force microscopy. *Phys. Rev. B* **71,** 155418 (2005).

30. Ishigami, M., Chen, J. H., Cullen, W. G., Fuhrer, M. S. & Williams, E. D. Atomic structure of graphene on $SiO_2$. *Nano Lett.* **7**, 1643-1648 (2007).

**Acknowledgement:**

We thank Qiuzi Li, E. H. Hwang and S. Das Sarma for discussions. This work is supported by NSF Grant No. DMR 08-04976, and U.S. ONR MURI. The NSF UMD-MRSEC shared equipment facilities were used in this work.

**Competing financial interests:**

The authors have no competing financial interests to acknowledge.

**Figure captions:**

**Figure 1: Potassium deposition on graphene.** The main panel shows the low temperature (20 K) gate voltage dependence of the conductivity for the pristine device (black) and successive depositions of potassium (colored). The top left inset shows an optical microscope image of the monolayer graphene device used in this experiment, with a schematic of the measurement circuit. The scale bar is 3 $\mu m$. The right inset shows the carrier-density-dependent conductivity of the pristine graphene device. The red curve is a fit to equation (1), and the blue curve is a fit to the correlated charged impurity model (see text for fit parameters and discussion).

**Figure 2: Charged-impurity-correlation induced improvement in graphene mobility. a**, Temperature dependence of graphene conductivity after potassium doping to produce a shift of the minimum conductivity point $\Delta V_{g,min}$ = 62 V. Dashed lines are fits using equation (3). The inset shows the temperature dependence of the field effect mobility. **b**, Temperature dependence of field effect mobility for seven different potassium doping levels as measured by $\Delta V_{g,min}$.

**Figure 3: Carrier-density dependence of graphene conductivity at various temperatures for three different potassium doping levels. a,** $\Delta V_{g,min}$ = 78 V. The temperatures are 21.8, 42.5, 100, 116.5, 130.1, 146.3, 156.6, 162.6 K. **b**, $\Delta V_{g,min}$ = 41V. The temperatures are 19.4, 50.1, 94.9, 112.8, 126.8, 141.9, 158.5 K. **c,** $\Delta V_{g,min}$ = 10V. The temperatures are 20.7, 132.7, 141.9, 150.4, 162, 177.2 K. For each set the curves are

ordered from lowest to highest conductivity; the lowest and highest temperatures are also indicated in each panel. The dashed lines are fits to equation (3). The densities of potassium used as global fit parameters are shown for each panel.

**Figure 4: Fit parameters for data in Figures 2 and 3 to theory of correlated impurity scattering.** The main panel shows the correlation length $r_c$ as a function of temperature for the seven sets of data at different potassium densities reflected in the shift of the minimum conductivity point $\Delta V_{g,min}$ indicated in the legend. The inset shows in log-log scale $\Delta V_{g,min}$ dependence of the potassium densities $n_K$ obtained from the fits. The red dots correspond to the seven sets of temperature dependence data and the grey squares are for other potassium doping levels measured at base temperature only. The lines are theoretical predictions discussed in the text[9].

**Figure 5: Temperature dependence of the minimum conductivity.** Black circles are for pristine graphene and colored symbols are for various potassium densities given by the shift of minimum conductivity point $\Delta V_{g,min}$ indicated in the legend.

**Figure1:**

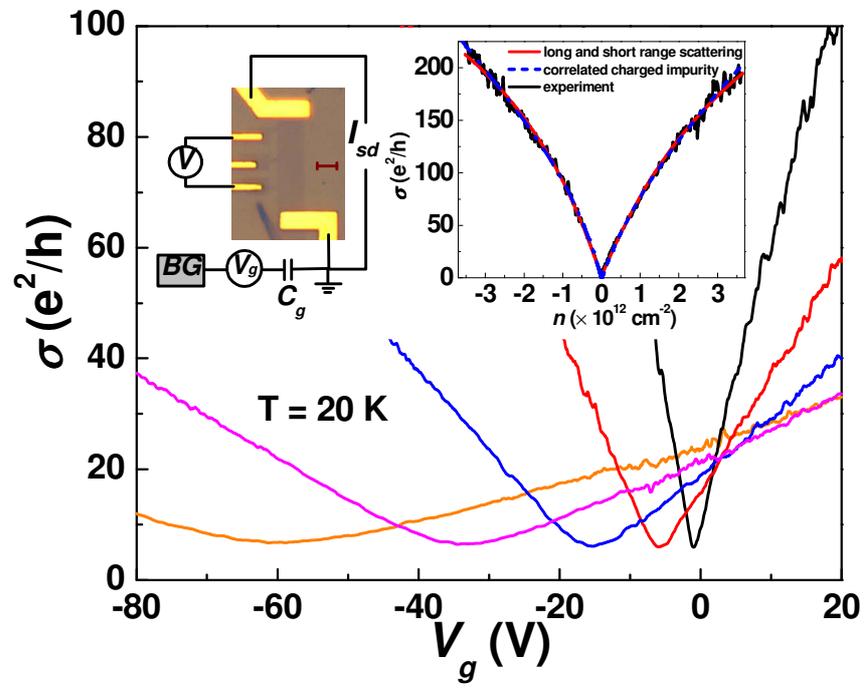

**Figure 2:**

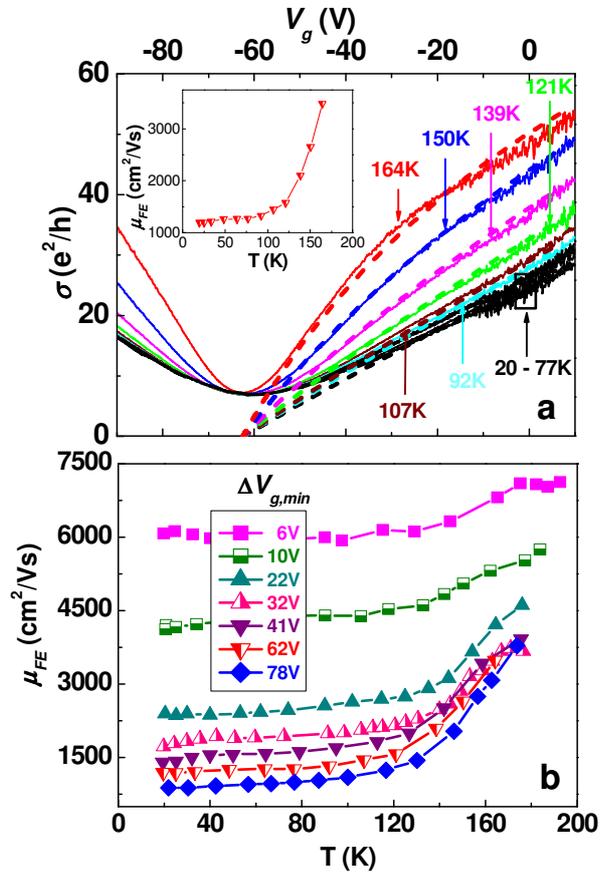

**Figure 3:**

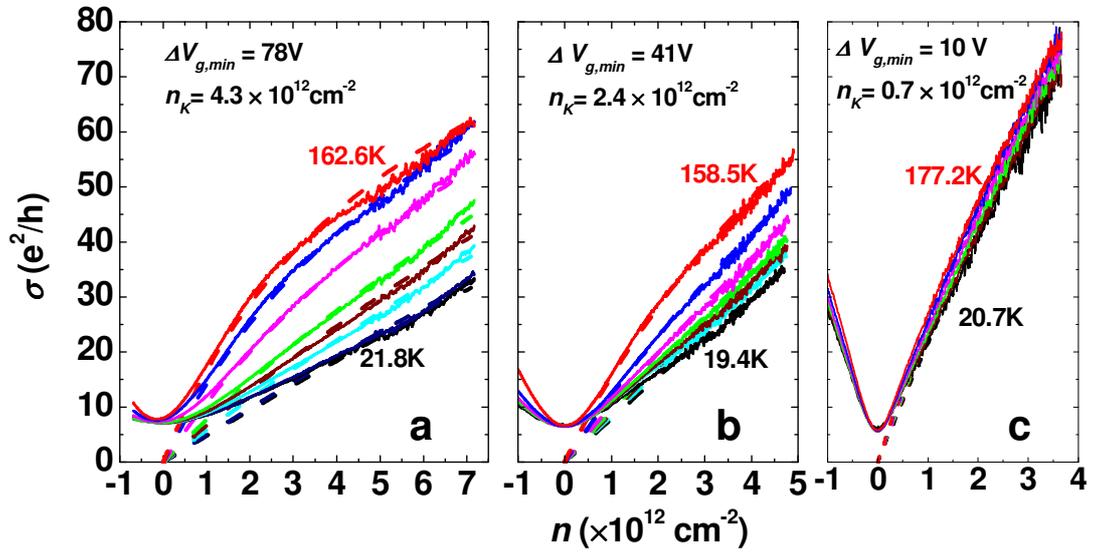

**Figure 4:**

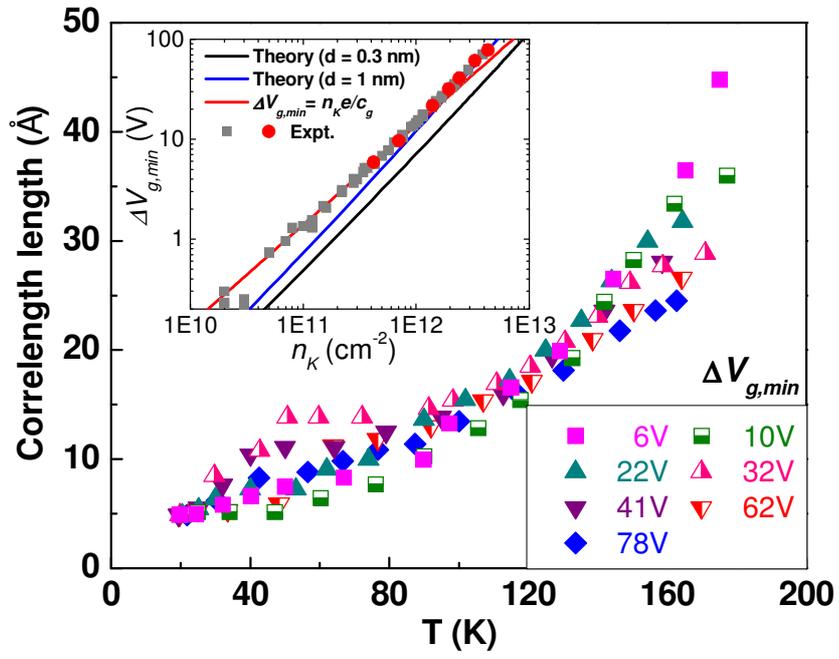

**Figure 5:**

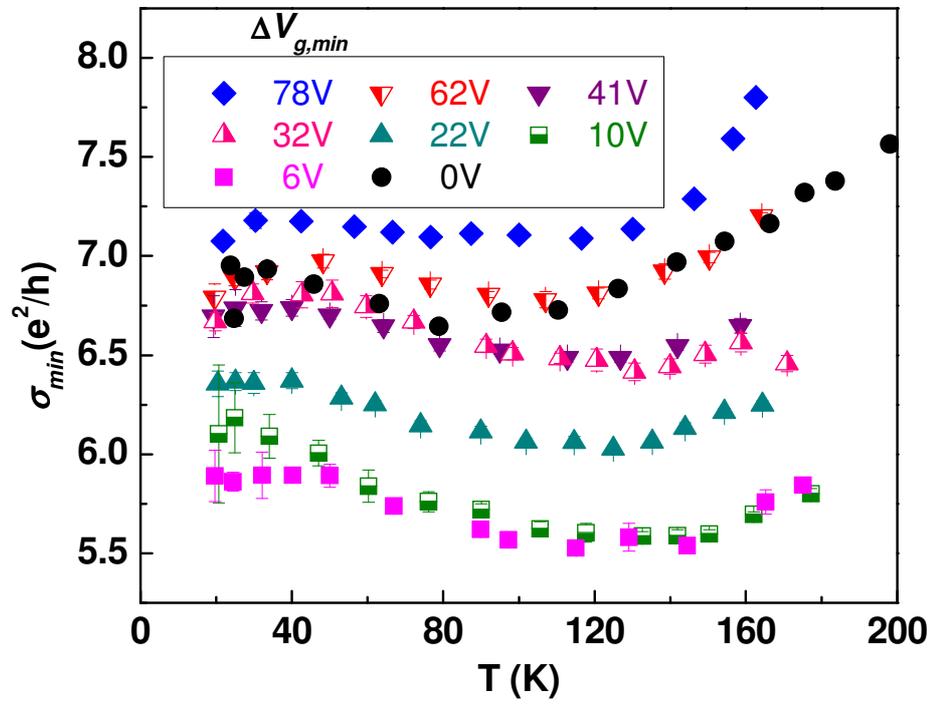

# Supplementary Information: Correlated charged impurity scattering in graphene

Jun Yan and Michael S. Fuhrer*

*Center for Nanophysics and Advanced Materials and Materials Research Science and Engineering Center, University of Maryland, College Park, MD 20742, USA*

* email: mfuhrer@umd.edu

**1. Fitting transport curves of the pristine graphene.**

Prior to potassium doping, we measure the conductivity of graphene sample at base temperature (~20 K) in ultrahigh vacuum (~$10^{-9}$ Torr). The electron density dependence of conductivity is fit with two methods.

The first method uses equation (1) assuming Matthiessen's rule that sums long range and short range scattering resistivities. The fit parameters are obtained by plotting $n/\sigma$ as a function of $n$:

$$\frac{n}{\sigma_{pristine}(n)} = \frac{1}{e\mu_L} + n\rho_S \qquad (S1).$$

This is linear; the intercept gives mobility $\mu_L$ and the slope gives $\rho_s$. Figure S1 shows a fit using this method.

The second method assumes that there are only long range scatters present but the scatters are correlated. The correlation is modeled with a structure factor equation (2) in the main text. The fit is obtained with a numerical integration as described in the Methods section. Figure S2 shows the fit of the same data as in Figure S1.

Both methods give overall good fits except very close to the charge neutral position where the physics is dominated by electron-hole puddles. As discussed in the Methods section of the manuscript, Method 1 can be mathematically viewed as an expansion of method 2 which explains why the two fits look very similar.

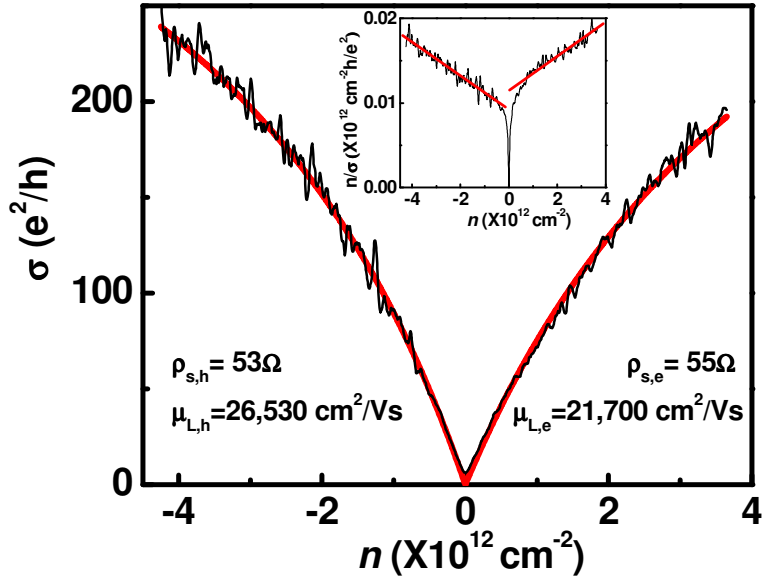

**Figure S1: Fitting pristine graphene transport using method 1.** The black curves are experimental measurements and the red curves are fits. The inset fit determines the fit parameters used and indicated in the main panel.

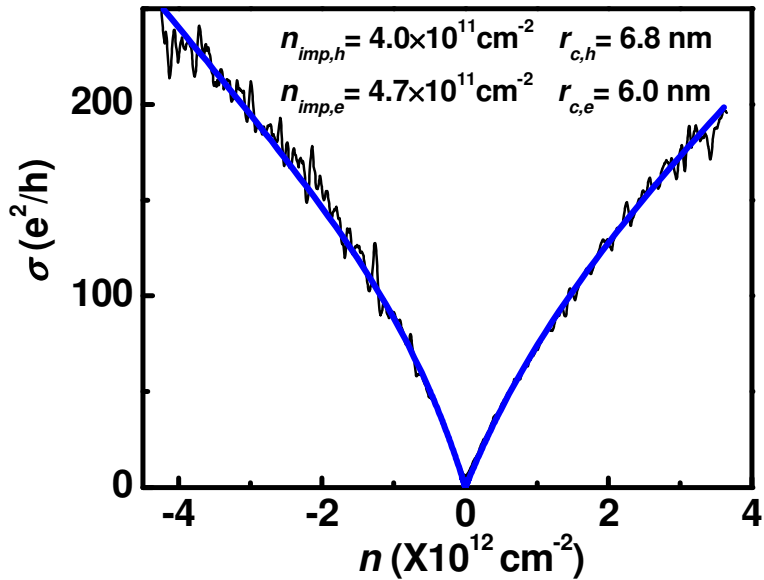

**Figure S2: Fitting pristine graphene transport using method 2.** The black curve is the same experimental data as in Figure S1. The blue curve is the fit with correlated charged impurities.

The two models are found to fit equally well all seven sets of pristine graphene transport prior to potassium doping. In Figure S3 we display the variations of these fit parameters as well as the slight drift of minimum conductivity position during repeated adsorption and removal of potassium atoms.

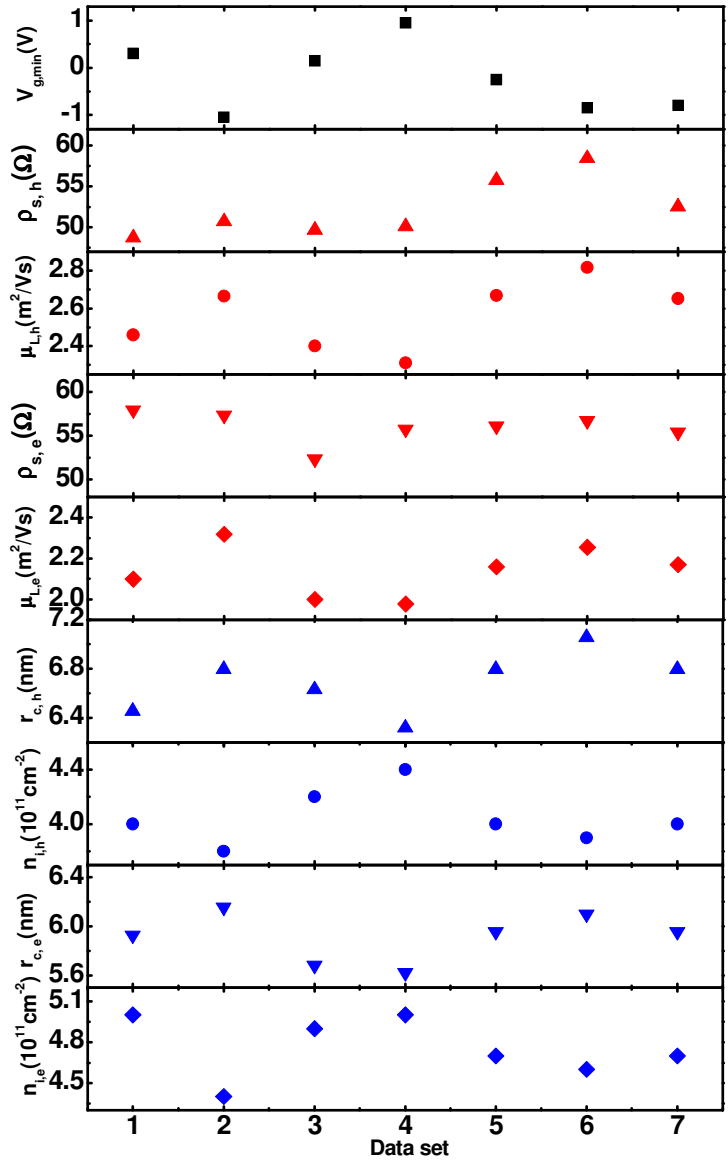

**Figure S3: Fit parameters for the pristine graphene.** All seven sets of data are shown. The red symbols are for method 1 and blue for method 2. The variations of the minimum conductivity positions are shown in the top panel.

## 2. Fitting transport traces of potassium doped graphene sample

We fit the temperature dependence of transport data with potassium doping to:

$$\sigma(n, n_K, T) = \left( \frac{1}{ne\mu_{L,e}} + \rho_{S,e} + \rho_{ph}(T) + \rho_K(n, n_K, r_c(T)) \right)^{-1} \quad (S3)$$

where $\mu_{L,e}$ and $\rho_{S,e}$ are from the fits to $\sigma_{pristine}(n)$ as discussed above for each set of data, $\rho_{ph}(T)$ is phonon contribution and $\rho_K(n, n_K, r_c(T))$ comes from scattering by correlated potassium ions.

We took in total 7 sets of data, four of which are shown in the main text (Fig.2a and Fig.3). The other 3 sets are shown below together with their fits.

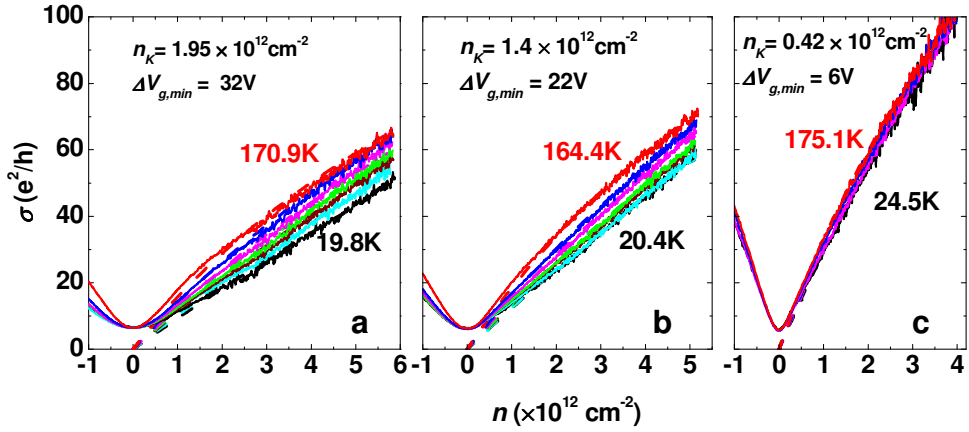

**Figure S4: Carrier-density dependence of graphene conductivity at various temperatures for three different potassium doping levels. a,** $\Delta V_{g,min}$ = 32 V. The temperatures are 19.8, 59.6, 120.6, 130.6, 139.9, 149.2, 170.9 K. **b,** $\Delta V_{g,min}$ = 22 V. The temperatures are 20.4, 53.1, 101.9, 114.6, 135.2, 143.9, 164.4 K. **c,** $\Delta V_{g,min}$ = 6 V The temperatures are 24.5, 144.5, 165.2, 175.1 K. For each set the lowest and highest temperatures are indicated in the caption. The dashed lines are fits to equation (3). The densities of potassium used as global fit parameters are shown for each panel.